\documentclass[runningheads]{llncs}
\usepackage[T1]{fontenc}
\usepackage{graphicx}

\usepackage{tabularx}
\usepackage{multirow}
\usepackage{array}
\usepackage{booktabs}
\usepackage{placeins}
\usepackage{float}


\begin{document}
\title{AI-Driven Analytics of Team-Teaching Talk: Acoustic Patterns across Experience, Cohorts and the Learning Design}
\titlerunning{AI-Driven Analytics of Team-Teaching Talk}

%
%

\author{Yuchen Liu\inst{1}\orcidID{0009-0008-3491-4261} \and
Roberto Martinez-Maldonado\inst{1}\orcidID{0000-0002-8375-1816} \and
Riordan Alfredo \inst{1}\orcidID{0000-0001-5440-6143} \and
Paola Mejia-Domenzain  \inst{2}\orcidID{0000-0003-1242-3134} \and
Dwi Rahayu\inst{1}\orcidID{0009-0008-5940-4911} \and
Sadia Nawaz\inst{1}\orcidID{0000-0002-3674-2108}}
\authorrunning{Y. Liu et al.}
%
\institute{
Monash University, Australia \\
\email{\{yuchen.liu1, Roberto.MartinezMaldonado, riordan.alfredo1, Dwi.Rahayu, sadia.nawaz\}@monash.edu}
\and
EPFL, Switzerland \\
\email{paola.mejia@epfl.ch}
}

\maketitle              
\vspace{-15pt}
\begin{abstract}
As classroom cohorts expand, team teaching is increasingly used to integrate the expertise and pedagogical perspectives of multiple teachers. Yet, there is limited empirical understanding of how team teaching unfolds in practice, particularly regarding differences in teachers’ contributions across experience levels, student cohorts, and learning task design.
Prior research on team teaching has largely relied on retrospective self-reports or small-scale observations, offering limited insight into the micro-level processes through which team teaching is enacted. Teacher talk offers a scalable lens on these processes. While research in individual teaching contexts shows that acoustic features of speech (e.g., voice quality, intonation, and loudness) can shape student learning, evidence from team-teaching settings remains scarce. Moreover, capturing such features through manual observation or transcription is especially challenging in team-teaching classrooms, where multiple teachers speak across extended sessions and spatial locations, limiting scalability without automation.
Grounded in spatial pedagogy theory and team-teaching research, this paper presents an AI-based speech processing approach to analyse classroom talk in team-teaching settings. We analysed 36 recorded undergraduate and postgraduate sessions involving 12 teachers. Spatial pedagogy behaviours were coded and acoustic features extracted to examine variation across teachers' experience, student cohorts, and the learning task design. The results reveal systematic differences, most notably in loudness dynamics: high-experience teachers, undergraduate classes and collaborative learning tasks exhibited greater loudness variation, suggesting more frequent modulation of volume to foreground key information and support classroom interaction and engagement.

\keywords{Spatial Pedagogy \and  Acoustic Analysis  \and Classroom Analytics}
\end{abstract}
\vspace{-20pt}
\section{Introduction}

As classroom cohorts expand, team teaching is increasingly adopted to enrich learning by integrating teachers’ complementary pedagogical perspectives and instructional practices \cite{marien2023teaching}. Grounded in a socio-constructivist view, it supports the co-construction of knowledge through interaction and dialogue, making ways of thinking, reasoning, and collaboration visible to learners \cite{Palincsar1998}. Prior work suggests that student learning benefits when such instructional practices are articulated coherently through interaction and dialogue \cite{dang2022academics,de2025studying}.

Yet, evidence on team teaching remains mixed. Recent empirical work suggests that differences between team teaching and solo teaching may be modest \cite{de2026team}, highlighting a gap between theorised pedagogical benefits and limited empirical understanding of how team teaching is enacted in practice and shaped by contextual conditions such as teachers’ experience, student cohorts, and learning task design \cite{de2026team}. Moreover, previous research has been methodologically constrained, relying largely on observational or self-report approaches, such as questionnaires assessing collaboration, shared responsibility, or teachers’ perceptions, which provide only indirect access to classroom enactment \cite{decuyper2023collaboration,decuyper2024complexity,baeten2014student,de2025studying}. Consequently, the micro-level processes through which team teaching unfolds, often beyond teachers’ explicit awareness, remain insufficiently understood \cite{Liu2025TeamTeaching,de2025studying}.

One theoretical approach for examining how teaching is enacted in classroom space is spatial pedagogy \cite{lim2012spatial}. This posits that classroom spaces acquire pedagogical meaning through the discourse enacted at particular physical classroom areas, as well as through teachers’ positioning and proximity to students and learning resources. Lim et al. \cite{lim2012spatial} distinguish four classroom space types, \textit{Authoritative}, \textit{Interactional}, \textit{Supervisory}, and \textit{Personal}, which can be characterised through observable \textit{spatial pedagogy behaviours} to study the micro-level enactment of teaching \cite{Liu2025TeamTeaching}. Although spatial pedagogy has primarily been applied to small-scale studies of solo teaching, it is  relevant to team-teaching contexts, where multiple teachers simultaneously share and negotiate classroom space \cite{baeten2014student}. However, spatial positioning alone is insufficient to fully capture enactment, as pedagogical meaning is also shaped by how teachers speak within these spaces.

Building on this perspective, Wu \cite{wu2025spatial} demonstrated that incorporating teacher talk into analyses of spatial pedagogy can enable finer-grained distinctions among instructional approaches. However, this work relied on discourse-level descriptions, leaving the acoustic realisation of teachers’ talk largely unexplored. This is a significant limitation, as ways of speaking, such as voice quality and loudness variation, can reshape, and at times counteract, the pedagogical meaning of what is said \cite{hamalainen2018s}. While prior research in solo teaching contexts shows that such acoustic features influence student comprehension and attention \cite{hamalainen2018s}, they remain largely unexamined in team-teaching settings.

Beyond spatial positioning, observational research suggests that teacher talk varies systematically with contextual factors \cite{wang2024artificial}. For instance, teacher experience shapes talk patterns, with less experienced teachers relying on more routinised routines and more experienced teachers exhibiting greater flexibility and variation \cite{tong2024exploring}. Student cohort characteristics also matter, as instruction at higher educational levels tends to involve more complex linguistic constructions and less reliance on common vocabulary \cite{csomay2007corpus}. Learning task design further influences whether teachers adopt facilitative roles in collaborative tasks or act as direct instructors in individual tasks \cite{shinde2022effectiveness}. However, this work has focused primarily on variation in the content and organisation of teacher talk, rather than on how talk is realised acoustically. Moreover, because this evidence is largely derived from solo-teacher classrooms, it remains unclear whether, and how, teacher talk in team-teaching settings varies with teacher experience, cohort characteristics and task design. Given that teacher talk is a key medium through which team teaching is enacted \cite{bonacina2025interactional}, there is a need to examine its acoustic features in relation to such contextual factors.

To address this gap, this paper presents an AI-based speech processing approach to analyse how teams of teachers talk in the classroom. We analysed 36 recorded team-teaching sessions from an undergraduate and postgraduate database course, combining coded spatial pedagogy behaviours with AI-based extraction of acoustic features from teacher talk. We examine how enactment varies across teacher experience, student cohorts, and learning task design, offering new evidence on team teaching from an acoustic and spatial perspective.

\vspace{-10pt}
\section{Background and Related Works}
\vspace{-3pt}
\subsection{Foundations of Team Teaching}
\vspace{-3pt}
Team teaching involves two or more teachers collaboratively delivering instruciton within the same classroom \cite{baeten2014student}. This approach has been associated with several pedagogical benefits. For students, team teaching can expose them to diverse instructional strategies and disciplinary perspectives, potentially enriching understanding and supporting active engagement \cite{kim2021does}. For teachers, it can foster collaboration and shared responsibility, creating opportunities for peer learning, professional development, and ongoing pedagogical reflection \cite{baeten2014student}.

Despite these theorised advantages, recent research emphasises that team teaching is not a uniform instructional approach but encompasses a range of practices whose impact depends on how they are enacted in the classroom \cite{decuyper2024complexity}. Empirical evidence further suggests that the benefits of team teaching are often modest and highly context-dependent \cite{de2026team}. This sensitivity to implementation is particularly pronounced in higher education, where team teaching is frequently adopted as an ad hoc, context-specific strategy rather than as a systematically structured instructional approach \cite{alvarez2024team}. In turn, a central challenge is to capture and explain the micro-level enactment of team teaching in classrooms and how it varies across contexts. One way to address this challenge is to draw on spatial pedagogy as an observational lens for examining classroom practice \cite{lim2012spatial}.

\vspace{-10pt}
\subsection{Spatial Pedagogy}
\vspace{-3pt}
The term \textit{spatial pedagogy} has been coined to refer to the patterns of positioning and movement that teachers enact within classroom spaces, where different areas acquire distinct pedagogical meanings depending on the relative positions and distances between teachers, students, and classroom resources \cite{lim2012spatial}. Within this framework, four space types are distinguished in the classroom:

\begin{itemize}
    \item The \textbf{supervisory space} refers to moments when teachers circulate around the classroom to monitor students’ work, maintaining a distance beyond students’ personal space and without engaging in direct interaction.
    \item The \textbf{authoritative space} refers to situations in which teachers position themselves at the front centre of the classroom—typically near the teacher’s desk and furthest from students—to deliver instruction, provide explanations, and manage whole-class activities.
    \item The \textbf{interactional space} refers to instances where teachers move among students, positioning themselves within students’ personal space to initiate dialogue, offer guidance or provide feedback.
    \item The \textbf{personal space} refers to periods when teachers remain largely stationary and distant from students—often behind the teacher’s desk or a similar resource—to organise materials, complete administrative tasks, or prepare subsequent lesson activities.
\end{itemize}

\vspace{-3pt}
Spatial pedagogy captures how positioning and movement organise classroom activity, but because the same spatial configuration can serve different instructional purposes, understanding team teaching in shared spaces also requires analysing how teachers’ talk enacts pedagogical meaning in situ.

\vspace{-10pt}
\subsection{AI and Analytics Approaches to Teachers' Talk}
\vspace{-3pt}
Teachers’ talk plays a central role in shaping instructional processes and students’ learning opportunities. Yet, traditional approaches rely heavily on manual transcription and human coding, making analysis costly, time-consuming, and difficult to scale across lessons and classrooms \cite{schlotterbeck2022teacher}. These constraints have motivated the use of AI-based speech analysis methods that enable the automated extraction of classroom-relevant information from audio recordings, supporting large-scale analyses beyond what manual approaches can achieve \cite{schlotterbeck2022teacher}.

Most existing AI-driven research on classroom speech has focused on discourse-level features. For example, prior work has automated the identification of teacher questioning and broader discourse features from classroom audio \cite{huang2020neural,jensen2021deep}. Beyond questioning, tools such as TalkMoves operationalize accountable-talk moves to support productive and equitable discussion \cite{suresh2021using}, and discourse indicators have been linked to external measures of teaching quality \cite{boyle2024semantic}. In contrast, research in speech science and education suggests that acoustic features convey pedagogically and socially meaningful information that can reinforce—or even counteract—the interpretation of spoken content \cite{hamalainen2018s}. Prior work demonstrated, for example, that teachers’ prosodic cues can shape students’ interpretations of evaluative stance and answer adequacy during question–answer interactions \cite{sikveland2021teachers}, and that paraverbal features can signal supportive or undermining discourse with consequences for students’ motivation, identity, and sense of belonging \cite{hunkins2022beautiful}. Similar studies further suggest that prosodic analysis can reveal how teachers adapt vocal delivery across instructional contexts and talk patterns in ways that transcript-based analyses alone cannot capture \cite{hamalainen2018s}. However, this body of work has largely focused on solo teacher classrooms. To date, no research has applied acoustic analysis to team-teaching settings, particularly to examine how teachers’ talk varies with contextual factors such as teacher experience, cohort characteristics, and learning task design.

\vspace{-10pt}

\subsection{Research Questions}
\vspace{-2pt}
Based on the gaps identified in previous research, the present study investigates the following research questions:

\begin{itemize}
  \item \textbf{RQ1:} To what extent can acoustic analysis reveal differences between high-experience and low-experience teachers in team teaching classrooms?
  
  \item \textbf{RQ2:} To what extent can acoustic analysis reveal differences in teacher talk when engaging different student cohort characteristics (i.e., undergrad versus postgrad) in team teaching classrooms?
  
  \item \textbf{RQ3:} To what extent can acoustic analysis reveal differences in teacher talk when supporting students in different learning task design (i.e., individual versus collaborative) in team teaching classrooms?
\end{itemize}

\vspace{-20pt}
\section{Methods}
\vspace{-5pt}
\subsection{Context}

\vspace{-5pt}
The study was conducted in an undergraduate (UG) and postgraduate (PG) Database course at Monash University. Course content and structure were identical across levels, enabling comparability. Classes were 2 hours long and delivered via team teaching, with three instructors jointly leading each session. Over four consecutive weeks, 36 sessions were observed (six UG and three PG per week). All instructors wore wireless headset microphones (\textit{Shure PGA31}) for audio capture; in two sessions, a microphone placement issue rendered one instructor’s audio unusable. Consequently, analyses included 34 sessions with complete recordings. Instructors rotated in triads as scheduled, with no research-driven modifications, and followed a shared weekly lesson plan. Within sessions, classroom activities were annotated as either individual tasks (e.g., students working independently) or collaborative tasks (e.g., collaborative activities involving peer discussion). A total of 62 classroom activities were annotated, including 44 individual tasks and 18 collaborative tasks.

\vspace{-10pt}
\subsection{Participants}
\vspace{-5pt}
Twelve teachers took part in the study. Their prior teaching experience ranged from zero to nine years. Based on a median split of teaching experience, participants were grouped into two categories: low-experience ($\leq 2$ years, $N = 6$) and high-experience ($> 2$ years, $N = 6$). In accordance with the approved ethics protocol (Protocol No. 28905, Monash University), no further personal data were collected. All participants provided informed consent.

\vspace{-10pt}
\subsection{Coding Scheme}
\vspace{-5pt}
Spatial pedagogy behaviours were systematically documented through structured classroom observation, using a coding scheme adapted from Alfredo et al.,\cite{alfredo2025teamteachingviz}. To ensure coding consistency, two observers first completed a briefing session to align their understanding of each spatial pedagogy behaviour. They then independently coded one pilot session. Inter-rater reliability on the pilot data reached a Cohen’s Kappa of 0.75, indicating good agreement. Discrepancies were then discussed to reach consensus and informed refinements and clarifications to the codebook. The remaining sessions were subsequently coded using the finalised codebook. The scheme integrates concepts from spatial pedagogy \cite{lim2012spatial} and team teaching \cite{cook1995co}. Specifically, the original scheme was informed by Lim et al.’s \cite{lim2012spatial}  four space types (\textit{Authoritative}, \textit{Interactional}, \textit{Supervisory}, and \textit{Personal}). To better capture talk in team teaching classrooms, the broad \textit{Interactional} type was refined into two talk-oriented categories: \textit{Interaction} (teacher-student interaction) and \textit{Collaboration} (teacher-teacher interaction). Furthermore, the present study focuses on teachers’ ways of speaking. Spatial pedagogy behaviours associated with \textit{Supervisory} and \textit{Personal} space types were not retained in the analytical scheme, as these behaviours typically involve little or no sustained speech. The final coding scheme distinguished six spatial pedagogy behaviours organised into three broader categories (see Table~\ref{tab:behaviours_codebook}).

\begin{table}[!ht]
    \centering
    \caption{Behaviour Coding Scheme.}
    \scriptsize
    \begin{tabularx}{\textwidth}{>{\raggedright\arraybackslash}p{2.2cm} >{\raggedright\arraybackslash}p{3.2cm} >{\raggedright\arraybackslash}p{2cm} X}
        \hline
        \textbf{Behavioural Category} & \textbf{Category Definition} & \textbf{Spatial Pedagogy Behaviour} & \textbf{Behaviour Definition} \\
        \hline
        \multirow{1}{=}{Instruction} & \multirow{1}{=}{Teachers directly teach students} 
        & Lecturing & A teacher delivers instructional material to the class. \\
        \hline
        \multirow{3}{=}{Interaction} & \multirow{3}{=}{Teachers interact with students}
        & Ts-S interaction & Multiple teachers interact with one student. \\
        & & T-S interaction & A teacher communicates one-to-one with an individual student. \\
        & & T-Ss interaction & A teacher offers guidance or explanation directed at a group of students. \\
        \hline
        \multirow{2}{=}{Collaboration} & \multirow{2}{=}{Teachers collaborate and interact with each other during the session}
        & Assisting & One teacher provides help or support to another. \\
        & & T-T interaction & Teachers engage in communication with each other. \\
        \hline
    \end{tabularx}
    \label{tab:behaviours_codebook}
\end{table}

\vspace{-10pt}
\subsection{Data Preprocessing}
\vspace{-5pt}

\begin{figure}[!ht]
    \centering
    \includegraphics[width=\linewidth]{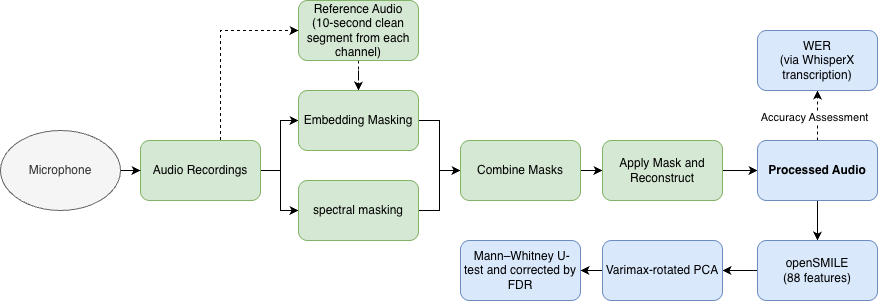}
    \caption{Overview of the audio preprocessing and analysis workflow. The eclipse represents the sensor. The green rectangles represent the data processing. The blue rectangles represent the processed data.}
    \label{fig:audio_preprocess}
\end{figure}

Due to multi-channel headset recordings in the team-teaching setting, substantial cross-talk was present. To mitigate this, we applied a weighted fusion of spectral masking \cite{mandel2014analysis} and ECAPA-TDNN speaker-embedding–based masking \cite{desplanques2020ecapa} to reduce cross-channel interference (Figure~\ref{fig:audio_preprocess}). To verify suitability for downstream analysis, we assessed transcription accuracy using WhisperX \cite{bain2023whisperx} and computed word error rate (WER) \cite{wang2003word}. The WER decreased from $158\%$ before cross-talk mitigation to $21\%$ after processing, which is considered acceptable \cite{munteanu2006measuring}. The processed audio was then used for acoustic feature extraction and statistical analyses.

\vspace{-10pt}
\subsection{Acoustic Analysis}
\vspace{-5pt}
We extracted 88 acoustic features using openSMILE \cite{eyben2010opensmile} from fixed 1-second audio segments in the processed audio files; this window length follows prior multimodal methodology for capturing moment-to-moment changes \cite{buckingham2019multimodal}.To ensure comparability across features with different scales, all acoustic features were standardised using z-score normalisation prior to dimensionality reduction. To reduce feature dimensionality and identify interpretable patterns, we performed principal component analysis (PCA) on the standardised acoustic features. Varimax rotation was applied to improve component interpretability. This procedure identified five rotated components (RC) that explained $58\%$ of the cumulative variance (see Table \ref{tab:pca_results}). 

\begin{table}[!htbp]
    \centering
    \caption{Results of the Principal Component Analysis (Varimax-rotated PCA).}
    \scriptsize
    \renewcommand{\arraystretch}{1.0}
    \begin{tabularx}{\textwidth}{
      >{\centering\arraybackslash}X
      >{\raggedright\arraybackslash}p{4.8cm}
      >{\raggedright\arraybackslash}X
      >{\raggedleft\arraybackslash}p{1.4cm}
      >{\raggedleft\arraybackslash}p{1.8cm}
    }
    \toprule
    \textbf{Component} &
    \textbf{Main parameters} &
    \textbf{Perceptual correspondence} &
    \textbf{Loading} &
    \textbf{Variance explained} \\
    \midrule
    
    \multirow{5}{*}{RC1}
    & Alpha Ratio (Voiced, mean)        & \multirow{5}{=}{Spectral slope and formants} & 0.22  & \multirow{5}{*}{39.76\%} \\
    & Hammarberg Index (Voiced, mean)   && -0.22 & \\
    & F2 (mean)                         && -0.20 & \\
    & F3 (mean)                         && -0.20 & \\
    & F1 (mean)                         && -0.20 & \\
    \midrule
    
    \multirow{4}{*}{RC2}
    & HNR (mean)                        & \multirow{4}{=}{Voice quality \& spectral structure} & -0.35 & \multirow{4}{*}{7.31\%} \\
    & Slope 0--500 (Voiced, mean)       && -0.33 & \\
    & Mfcc4 (Voiced, mean)              && 0.31  & \\
    & Mfcc2 (Voiced, mean)              && 0.29  & \\
    \midrule
    
    \multirow{5}{*}{RC3}
    & Spectral Flux (mean)              & \multirow{5}{=}{Loudness dynamics} & 0.33 & \multirow{5}{*}{5.81\%} \\
    & Spectral Flux (Voiced, mean)      && 0.31 & \\
    & Loudness Rising Slope             && 0.29 & \\
    & Loudness Rising Slope (sd)        && 0.29 & \\
    & Loudness Range                    && 0.28 & \\
    \midrule
    
    \multirow{5}{*}{RC4}
    & Hammarberg Index (Unvoiced, mean) & \multirow{5}{=}{Unvoiced spectral characteristics} & -0.50 & \multirow{5}{*}{2.73\%} \\
    & Alpha Ratio (Unvoiced, mean)      && 0.49  & \\
    & Slope 0--500 (Unvoiced, mean)     && 0.45  & \\
    & Mfcc4 (mean)                      && -0.33 & \\
    & Mfcc3 (mean)                      && -0.21 & \\
    \midrule
    
    \multirow{5}{*}{RC5}
    & Loudness (sd)                     & \multirow{5}{=}{Loudness \& spectrum variation} & 0.35  & \multirow{5}{*}{2.48\%} \\
    & Spectral Flux (sd)                &                                                & 0.34  & \\
    & Loudness Peaks Per Sec            &                                                & -0.29 & \\
    & F0 Rising Slope (sd)              &                                                & -0.25 & \\
    & F0 Falling Slope (sd)             &                                                & -0.22 & \\
    \bottomrule
    \end{tabularx}
    \label{tab:pca_results}
\end{table}
\vspace{-5pt}

Component scores were then used for group comparisons across the three spatial pedagogy behavioural categories (Instruction, Interaction, and Collaboration) and the three factors (teachers' experience, student cohort characteristics and learning task design). Group differences to address RQ1-3 were assessed using  Mann–Whitney U-test and corrected by false discovery rate (FDR).

\vspace{-10pt}

\section{Results}
\vspace{-5pt}
\subsection{RQ1: Teachers' Experience and Teachers' Talk}

\begin{figure}[!ht]
    \centering
    \includegraphics[width=0.98\linewidth]{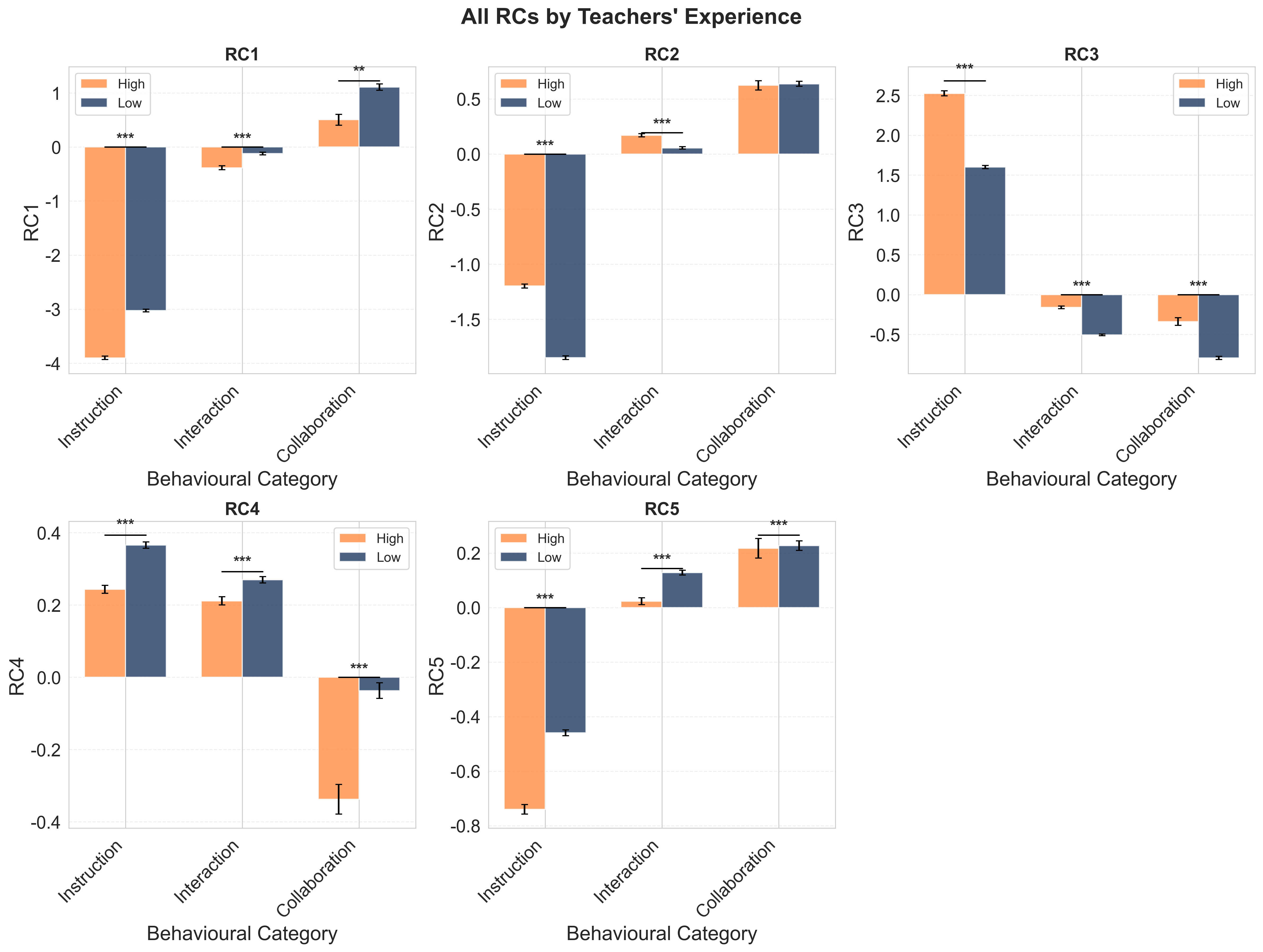}
    \caption{Comparison of all rotated principal components (RC1-RC5) by teachers' experience across three spatial pedagogy behavioural categories. Bar plots show mean values with $95\%$ confidence intervals for high-experience (orange) and low-experience (dark blue) groups. Significant differences were determined by the Mann–Whitney U-test and corrected by FDR ($\ast\mathrm{FDR} < 0.05$, $\ast\ast\mathrm{FDR} < 0.01$, $\ast\ast\ast\mathrm{FDR} < 0.001$).}
    \label{fig:experience_result}
\end{figure}

Significant differences in rotated components (RC) were observed between high- and low-experience teachers across all spatial pedagogy behavioural categories (see Figure~\ref{fig:experience_result}). High-experience teachers exhibited higher values in RC2 and RC3. Higher RC2 values indicated differences in voice quality and spectral structure. Higher RC3 values were associated with greater loudness dynamics, characterised by increased spectral flux, steeper loudness rising slopes, and a wider loudness range. In contrast, low-experience teachers showed higher values in RC1, RC4, and RC5. Higher RC1 values corresponded to a higher alpha ratio and lower mean formant frequencies (F1–F3). Higher RC4 values reflected increased modulation of unvoiced spectral characteristics. Higher RC5 values indicated greater overall loudness and spectral variability, with lower variability in F0 slope.

\vspace{-5pt}
\subsection{RQ2: Student Cohort Characteristics and Teachers' Talk}

\begin{figure}[!ht]
    \centering
    \includegraphics[width=0.98\linewidth]{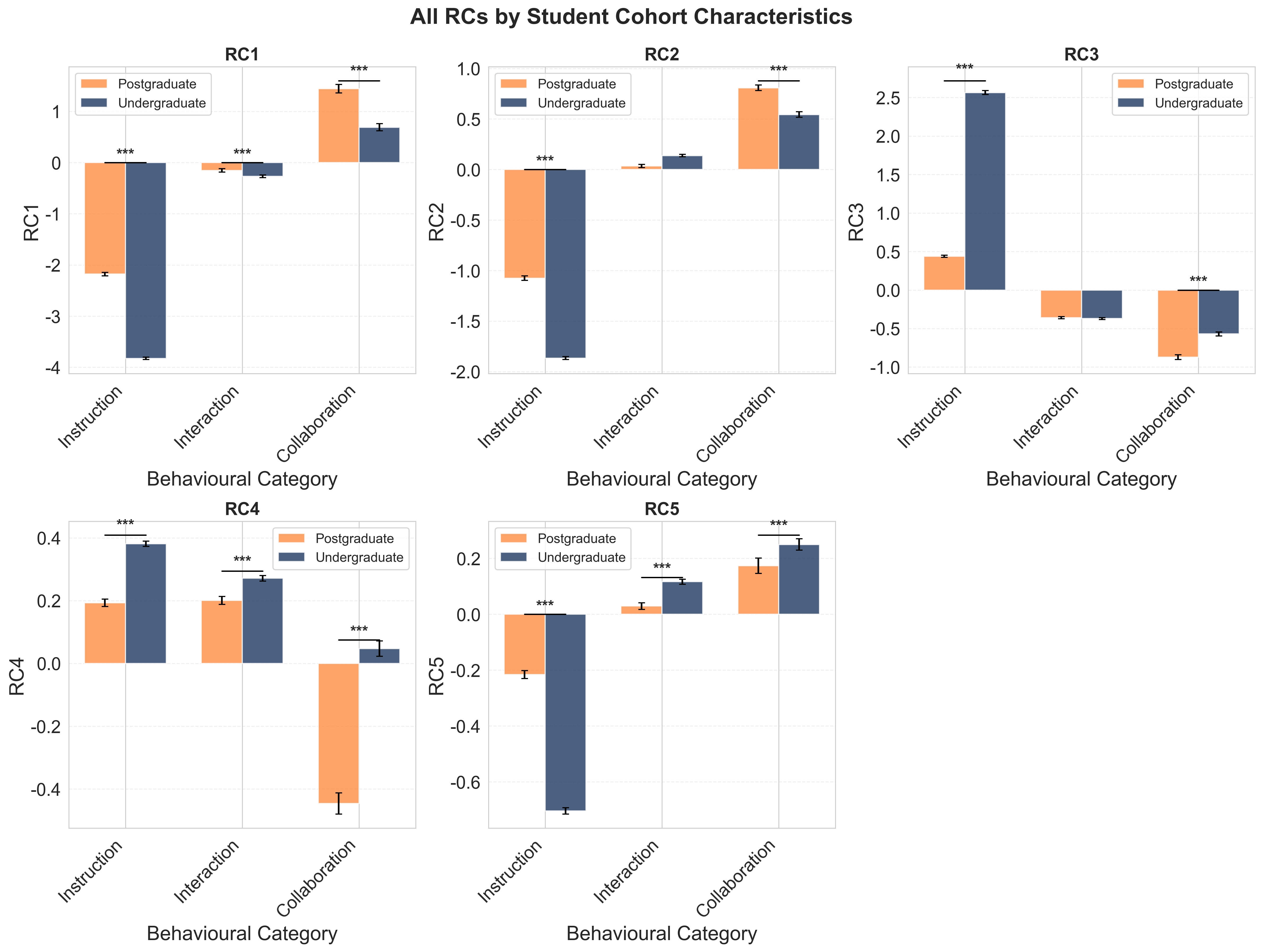}
    \vspace{-8pt}
    \caption{Comparison of all rotated principal components (RC1-RC5) by student cohort characteristics across three spatial pedagogy behavioural categories. Bar plots show mean values with $95\%$ confidence intervals for postgraduate-teaching (orange) and undergraduate-teaching (dark blue) groups. Significant differences were determined by the Mann–Whitney U-test and corrected by FDR ($\ast\mathrm{FDR} < 0.05$, $\ast\ast\mathrm{FDR} < 0.01$, $\ast\ast\ast\mathrm{FDR} < 0.001$).}
    \label{fig:degree_result}
    \vspace{-4pt}
\end{figure}

Across all spatial pedagogy behavioural categories, teaching postgraduate students showed significant differences in rotated components (RC) compared to teaching undergraduate students (see Figure~\ref{fig:degree_result}). The postgraduate-teaching group exhibited higher values in RC1 and RC2, whereas the undergraduate-teaching group showed higher values in RC3 and RC4. RC5 differences varied across categories, with higher values observed in the undergraduate-teaching group during Interaction and Collaboration, and higher values in the postgraduate-teaching group during Instruction.

\vspace{-10pt}
\subsection{RQ3: Learning Task Design and Teachers' Talk}

\begin{figure}[!ht]
    \centering
    \includegraphics[width=0.98\linewidth]{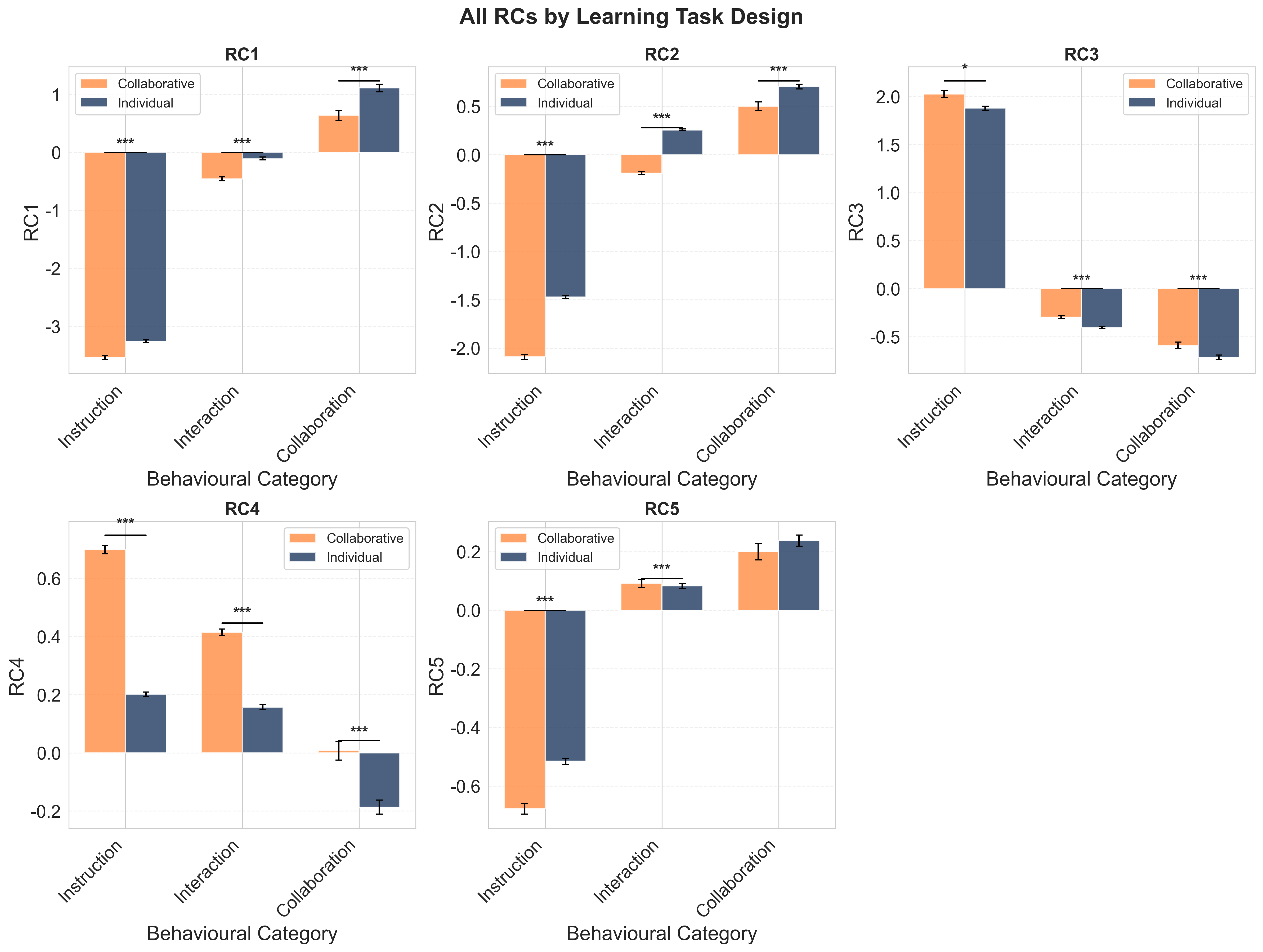}
    \vspace{-8pt}
    \caption{Comparison of all rotated principal components (RC1-RC5) by learning task design across three spatial pedagogy behavioural categories. Bar plots show mean values with $95\%$ confidence intervals for collaborative (orange) and individual (dark blue) tasks. Significant differences were determined by the Mann–Whitney U-test and corrected by FDR ($\ast\mathrm{FDR} < 0.05$, $\ast\ast\mathrm{FDR} < 0.01$, $\ast\ast\ast\mathrm{FDR} < 0.001$).}
    \label{fig:task_result}
\end{figure}

\vspace{-3pt}
Across all spatial pedagogy behavioural categories, individual tasks showed significant differences in rotated components (RC) compared to collaborative tasks (see Figure~\ref{fig:task_result}). Collaborative tasks exhibited higher values in RC3 and RC4, whereas individual tasks showed higher values in RC1 and RC2. RC5 differences varied across categories, with higher values observed in individual tasks during Instruction, and higher values in collaborative tasks during Interaction.

\FloatBarrier

\vspace{-10pt}
\section{Discussion}
\vspace{-6pt}
\subsection{Discussion by Research Question}
\vspace{-3pt}

Overall, the findings illustrate that teachers’ talk differs systematically across teachers' experience, student cohort characteristics and learning task design, and that these differences are evident across spatial pedagogy behavioural categories. \textit{Regarding RQ1}, high-experience teachers showed a more regulated yet dynamically modulated vocal profile (lower RC5 and higher RC3), whereas low-experience teachers exhibited a more variable profile with reduced loudness-related temporal dynamics and stronger unvoiced spectral modulation (higher RC5, lower RC3 and higher RC4). This suggests that greater teaching experience is associated with more regulated yet dynamically modulated teacher talk, extending prior discourse-based evidence that high-experience teachers show greater flexibility \cite{tong2024exploring} by demonstrating that such differences are also observable at the acoustic level.

\textit{Regarding RQ2}, when teaching postgraduate students, teachers in their teams demonstrated a more regulated and restrained vocal profile, with a shift in spectral slope and formant-related configuration (higher RC1) and reduced modulation of unvoiced spectral features (lower RC4), while showing reduced loudness-related temporal dynamics (lower RC3).  In contrast, when teaching undergraduate students, teaching teams exhibited a more dynamically modulated vocal profile, characterised by stronger loudness-related temporal dynamics (higher RC3), stronger unvoiced spectral modulation (higher RC4), and greater loudness and spectral variability during Interaction and Collaboration (higher RC5). This may indicate that undergraduate and postgraduate teaching impose different communicative demands on teachers, potentially requiring more dynamic talk to organise and energise interaction at the undergraduate level, and more restrained delivery to support focused conceptual progression at the postgraduate level. This is consistent with prior findings that teacher talk shifts across student cohort characteristics \cite{csomay2007corpus}. \textit{Regarding RQ3}, when teaching collaborative tasks, teams exhibited a more dynamically modulated vocal profile (higher RC3 and higher RC4). This pattern may reflect a more interactionally responsive vocal style that could support coordination and engagement in collaborative activity. In contrast, when teaching individual tasks, teaching teams demonstrated a more restrained vocal profile (lower RC3 and lower RC4).

\vspace{-10pt}
\subsection{Implications for Research and Educational Practice}

\vspace{-3pt}
Our findings  have implications for both research and educational practice. For research, this work introduces methodological mechanisms for analysing team teaching that move beyond reliance on retrospective self-report and manual observation \cite{decuyper2023collaboration,decuyper2024complexity}. By combining spatial pedagogy coding with automated acoustic analysis, the approach enables scalable, fine-grained examination of what actually unfolds when multiple teachers share instruction, opening new avenues for studying spatial pedagogy \cite{wu2025spatial} and team teaching through classroom-level, automated evidence \cite{decuyper2023collaboration,decuyper2024complexity,baeten2014student,de2025studying}.

For educational practice, the findings suggest that professional development for team teaching should treat teacher voice as a pedagogical resource for coordinating classroom activity, rather than merely a channel for transmitting content. Deliberate modulation of loudness can foreground key information, signal transitions, and support engagement across lesson phases. Making such tacit vocal strategies explicit through targeted training may help less experienced teachers develop adaptive vocal repertoires for team-teaching contexts.

\vspace{-10pt}

\subsection{Limitations and Future Work}
\vspace{-3pt}

This study has several limitations that also point to promising directions for future research. First, the analysis was conducted within a single institutional and language context, which may limit generalisability. However, examining team teaching in an authentic classroom setting provides an important first step in identifying how acoustic patterns unfold during instruction. Future work should extend this approach to additional contexts and to other factors that may influence team teaching, such as class size \cite{de2026team} and teacher characteristics (e.g., teacher competencies, interpersonal and communication skills) \cite{de2025studying}.
Second, acoustic measures were derived from classroom audio without richer multimodal information (e.g., gaze, gesture, spatial positioning). Nonetheless, an audio-focused approach enables scalable analysis across extended sessions and multiple teachers, and the preprocessing pipeline demonstrates feasibility in a complex team-teaching environment. Incorporating multimodal data in future studies would support more nuanced interpretation of the pedagogical functions of acoustic variation and improve measurement robustness.
Finally, the unit of analysis was the individual teacher, which limits inferences about team-level coordination. Even so, this focus provides a necessary foundation for establishing whether experience-related acoustic differences are detectable in team-teaching classrooms. Future research can build on this by adopting the teaching team as the analytic unit and modelling coordination directly through integrated discourse and acoustic analyses.

\vspace{-10pt}
\section{Concluding Remarks}
\vspace{-5pt}
This study coded spatial pedagogy behaviours and extracted acoustic features to examine how their enactment varies across three contextual factors. The results reveal differences, most notably in loudness dynamics: high-experience teachers, undergraduate classes and collaborative learning tasks exhibited greater loudness variation, suggesting more frequent modulation of volume to foreground key information and support classroom interaction and engagement. These findings underscore that patterns of teacher talk offer important insights into how team teaching is enacted in the classroom across different contextual factors.

\vspace{-10pt}
%
%
%
\bibliographystyle{splncs04}
\bibliography{mybibliography}
\end{document}